# Magnetic moment versus tensor charge

Mustapha MEKHFI

International Center for Theoretical Physics Trieste, Italy

We express the baryon magnetic moments in terms of the baryon tensor charges, considering the quarks as relativistic interacting objects. Once tensor charges get measured accurately, the formula for the baryon magnetic moment will serve to extract precise information on the quark anomalous magnetic moment, the quark effective mass and the ratio of the quark constituent mass to the quark effective mass. The analogous formula for the baryon electric dipole moment is of no great use as it gets eventually sizable contributions from various CP- violating sources not necessary associated to the quark electric dipole moment.

12.39.Ki, 13.40.Em

I – Introduction

A model independent, field theoretical relation between baryon magnetic moment and the spin structure of the baryon including the tensor charge has been derived by the authors of reference [1]. The expression they obtain explained why quark model is a good approximation in describing baryon magnetic moments (m.m). To find the relation between baryon magnetic moment and the spin structure they adopt a relativistic quantum mechanical approach by writing the non interacting fermion field as a sum of the quark and the antiquark field. The quark-antiquark pair creation and annihilation also contribute to the magnetic moment once sea quark excitation mixing is included in baryon ground states. For this purpose they express the baryon state as a Fock space expansion

$$|B\rangle = c_0 |q^3\rangle + \sum_\alpha c_\alpha |q^3 q\bar{q}\rangle_\alpha + \cdots$$

then they show for illustration that in a constituent quark model with valence $q^3$ and sea $q^3 q\bar{q}$ mixing for instance pair creations only contribute a small amount to the magnetic moment of the proton ($-0.065 \mu_P$ with $\mu_P$ the nucleon magneton) .It is worth to note that the inclusion of the sea quarks through the Fock space is a tentative to include quark interactions into the scheme. In this paper we reconsider the problem using a standard approach in which the baryon is made of valence quarks only. Then show that the various potential interactions only add small contributions except the ones contributing to the quark anomalous magnetic moments (a.m.m) thus generalizing the formula first given in reference [1]. An analogous formula for the baryon electric dipole moment has also been derived . In the following we assume we have derived an effective lagrangian defined at the scale of the low-energy electric and magnetic moments after having integrated all unwanted fields .For the magnetic moment we will retain the following operators ( i= u, d, s )

$$m_i \bar{\psi}\psi , \ Q_i \bar{\psi}\gamma_\mu \psi A^\mu, \frac{1}{2} a_i \frac{Q_i}{m_i} \bar{\psi}\sigma_{\mu\nu}\psi F^{\mu\nu} \ldots$$

and for the electric dipole moment we may consider the operators of potential contributions

$$\theta m_i \bar{\psi}\gamma_5\psi, \frac{i}{2}d_i\bar{\psi}\sigma_{\mu\nu}\gamma_5\psi F^{\mu\nu}, \frac{i}{2}d_{ic}\bar{\psi}T_a\sigma_{\mu\nu}\gamma_5\psi G_a^{\mu\nu}, -\frac{1}{6}cf_{\alpha\beta\gamma}G_{\alpha\mu\rho}G_{\beta\nu}{}^{\rho}G_{\gamma\lambda\sigma}\varepsilon^{\mu\nu\lambda\sigma}....$$

The exact expressions of the above coefficients $m_i, Q_i, a_i, \theta, d_i, d_{ic}, c$ in terms of the primordial coefficients defined at high energy (above the chiral- symmetry- breaking scale for instance),do contain renormalization effects as well as the effects of the integrated fields (Higgs field, heavy quarks, supersymmetric particles etc).The exact expressions are however model dependant but this is not relevant to our analysis. The operators above are not all intrinsic to the quark but still they contribute globally to the low energy moments .The triple gluon operator $cf_{\alpha\beta\gamma}G_{\alpha\mu\rho}G_{\beta\nu}{}^{\rho}G_{\gamma\lambda\sigma}\varepsilon^{\mu\nu\lambda\sigma}$ [2] for instance couples to the neutron with possibly one gluon for each light quark in the neutron hence generates a neutron electric dipole moment (e.d.m) which is the result of the collective "cooperation" of three different quarks. Such effect could not be factorized in terms of individual quarks and hence could not be written in terms of the baryon tensor charge. This is not surprising as we already know that the tensor charge do not couple to gluons [3]. The chromoelectric dipole moment $\frac{i}{2}d_{ic}\bar{\psi}T_a\sigma_{\mu\nu}\gamma_5\psi G_a^{\mu\nu}$ where $T_a$ and $G_a^{\mu\nu}$ denote a generator, and the field- strength tensor, for color $SU(3)_3$, is another operator whose effect cannot be associated to individual quarks. It enters the process in the form of a correction to the t-channel gluon exchange between bound quarks, making the hadronic wave function CP-violating. The "exchange" magnetic moments which are a generic result in any interacting field theory are yet other contributions which do contribute a non-additive piece to the baryon magnetic moment as they may involve two quarks or more .In a pure S-wave case their contributions simplify and read

$$\vec{\mu}_z = -i\sum_{i\prec j} P_{ij}(\vec{\sigma}_i \times \vec{\sigma}_j)[(Q_i - Q_j)g_{ij}^{(1)} + (Q_i + Q_j)g(m_i - m_j)g_{ij}^{(2)}]$$

Where $g_{ij}$ is a function of the positions of quarks $i$ and $j$ but are independent of the quark masses and the quark charges. $P_{ij}$ exchange the flavor labels of particles $i$ and $j$.

We will show that exchange magnetic moments have small contributions to the baryon m.m while for the baryon e.d.m all contributions intrinsic or non-additive seem equally sizeable and very model dependant.

The magnetic moment $\vec{\mu}_N$ involves the operator $Q_i \bar{\psi}\gamma_\mu \psi A^\mu$ and is by definition

$$-\vec{\mu}_N \cdot \vec{B} = -\sum_{i,\bar{i}} Q_i \int dr^3 \bar{\psi}_i \gamma_\mu \psi_i A^\mu$$

Inserting the static potential $\vec{A} = \frac{1}{2}\vec{B} \times \vec{r}, A^0 = 0$ into the above equation we get

$$\vec{\mu}_N = \langle PS | \sum_{i,\bar{i}} \frac{Q_i}{2} \int dr^3 \vec{r} \times \bar{\psi}_i \vec{\gamma} \psi_i | PS \rangle \qquad \text{(Eq:1)}$$

$Q_i, i = u, d, s$ are the quark charges, $\psi_i (\psi_{\bar{i}})$ the constituent quark (antiquark) field and $|PS\rangle$ is the baryon ground state with momentum $P$ and spin polarization $S$. The contributions of the quarks to the baryon spin are encoded in the axial and tensor charges, respectively denoted $\Delta i$ and $\delta i$. The axial charge is $\Delta i = \Delta_i + \Delta_{\bar{i}}$ where $\Delta_i(\Delta_{\bar{i}})$ is related to the expectation value of the relativistic quark(antiquark) spin operator in the baryon

$$\langle PS | \int dx^3 \psi_i^\dagger \vec{\Sigma} \psi_i | PS \rangle = 2\Delta_i \vec{S}$$

and for a parton model in the infinite momentum frame $\Delta_i$ can be shown to be related to the helicity difference $\Delta_i = \int dx \left[ q_{i\uparrow}(x) - q_{i\downarrow}(x) \right]$ with $q_{i\uparrow}(x), (q_{i\downarrow}(x))$ the probability of finding a quark with fraction $x$ of the baryon momentum and polarization parallel (anti parallel) to the baryon spin. Similarly, the tensor charge is $\delta i = \delta_i - \delta_{\bar{i}}$ and $\delta_i$ is given by the formula

$$\langle PS | \int dx^3 \overline{\psi}_i \vec{\Sigma} \psi_i | PS \rangle = \vec{\delta}_i$$

and is related to the quark transversity distribution $q_{\to i}(x) - q_{\leftarrow i}(x)$ as a first moment $\delta_i = \int_0^1 dx [q_{\to i}(x) - q_{\leftarrow i}(x)]$ [4]. Similar expressions apply to the antiquark. Together, the unpolarised quark distribution (well known), the quark helicity distribution (known), and the transversity (unmeasured but calculated on the lattice, last calculations: $\delta i = 0.562 \pm 0.088$) provide a complete description of the quark spin. If quarks moved nonrelativistically in the nucleon, $\delta_i(x)$ and $\Delta_i(x)$ would be identical as only large components of the fermion field are leading in which case $\overline{\psi} = \psi \dagger \gamma^0 \simeq \psi \dagger$. Another way of seeing this is that rotations and Euclidean boosts commute and a series of boosts and rotations can convert a longitudinally polarized nucleon into a transversely polarized nucleon at infinite momentum. So the difference between the transversity and helicity distributions reflects the relativistic character of quark motion in the nucleon. To obtain the formula relating the baryon magnetic moment to the hadronic tensor we compute (Eq:1) using the free field current $\vec{j}_i = \overline{\psi}_i \vec{\gamma} \psi_i$ and assume that the ground state of the baryon has a vanishing non-relativistic orbital magnetic moment. To this end it is useful to decompose the quark current into two pieces using Gordon decomposition. The convection current part and the spin part contribute differently giving respectively

$$\frac{Q_i}{4(m+\bar{m})}(\Delta_i - \frac{\bar{m}}{m}\delta_i)$$

$$\frac{Q_i}{4\bar{m}}(\Delta_i + \frac{\bar{m}}{m}\delta_i)$$

Adding the quark contributions together to those of the antiquarks we get

$$\mu_N = \sum_{i=u,d,s} \frac{Q_i}{2\bar{m}_i}\left[\frac{1}{2(1+\frac{m_i}{\bar{m}_i})}(\Delta_i - \Delta_{\bar{i}} - \frac{\bar{m}_i}{m_i}\delta i) + \frac{1}{2}(\Delta_i - \Delta_{\bar{i}} + \frac{\bar{m}_i}{m_i}\delta i)\right] \quad \text{(Eq:2)}$$

With $\vec{\mu}_N = \langle P\uparrow | \vec{\mu}_N | P\uparrow \rangle$ and where the effective quark mass $\bar{m}_i$ is the average of the quark energy in the baryon ground state. Note that the magnetic moment involves the tensor charge $\delta i$ and the combination $\Delta_i - \Delta_{\bar{i}}$ which is not the axial charge $\Delta i$.

To obtain an analogous formula for the e.d.m of the baryon we consider the term $\frac{i}{2}d_i\bar{\psi}\sigma_{\mu\nu}\gamma_5\psi F^{\mu\nu}$ and get by definition of the e.d.m

$$H_I = -\vec{E}.\langle P\uparrow | \vec{d}_N | P\uparrow \rangle$$
$$= -i\langle P\uparrow | \sum_{i,\bar{i}} \int d_i \bar{\psi}_i \sigma^{\mu\nu}\gamma_5 \psi_i \partial_{\mu\nu} A_\nu dr^3 | P\uparrow \rangle$$

the expression of the nucleon e.d.m in terms of the $\delta_i + \delta_{\bar{i}}$. In effect

$$\vec{d}_N = \langle P\uparrow| \vec{d}_N | P\uparrow\rangle$$
$$= \langle P\uparrow| \sum_{i=u,d,s} d_i \int (\bar{\psi}_i \vec{\Sigma} \psi_i) dr^3 | P\uparrow\rangle$$
$$= \sum_{i,\bar{i}} d_i \vec{\delta}_i - d_{\bar{i}} \vec{\delta}_{\bar{i}}$$
$$= \sum_{i,\bar{i}} d_i (\vec{\delta}_i + \vec{\delta}_{\bar{i}})$$

In the above steps, the field strength has been set to correspond to a static electric field $A_\nu = (\vec{E}.\vec{r}, 0)$ and the identities $\sigma^{\mu\nu}\gamma_5 = \frac{i}{2}\varepsilon^{\mu\nu\rho\sigma}\sigma_{\rho\sigma}$, $\Sigma^i = \frac{1}{2}\varepsilon^{ijk}\sigma_{jk}$ has been used together with the minus sign added to the definition of the antiquark distribution to account for the charge conjugation oddness of the tensor field. For a baryon at rest and polarized along the third direction ($(\vec{\delta}_i)_3 = \delta_i$) we get

$$d_N = \sum_{i=u,d,s} d_i (\delta_i + \delta_{\bar{i}}) \qquad (Eq:3)$$

where $d_i$ is the e.d.m of the quark $i$. Note that the formula for the e.d.m involves the combination $\delta_i + \delta_{\bar{i}}$ and not the baryon tensor charge $\delta i$.

II – Adding anomalous magnetic moment of quarks

There is a term in the effective lagrangian which adds an extra (anomalous) piece to the intrinsic magnetic moment of the quark $\frac{1}{2} a_i \frac{Q_i}{m_i} \bar{\psi} \sigma_{\mu\nu} \psi F^{\mu\nu}$. Nonlinear chiral quark model [5] for instance can be used to estimate the order of magnitude of the anomalous contribution. In fact one would expect an a.m.m of order $\frac{m_i^2}{\Lambda_{CSB}^2} \approx 10\%$ ($m_i \approx 360 Mev$, $\Lambda_{CSB} \approx 1 Gev$) where $m_i$ is the constituent mass of the quark which is supposed to be the effect of chiral symmetry breaking, and $\Lambda_{CSB}$ is the chiral symmetry breaking scale. There are several theoretical and experimental studies indicating that quarks do have an a.m.m . To fit the measured magnetic moment of the baryon octet, it is found that the quarks must have a sizable a.m.m [6]. Bicudo et al [7] have shown in several effective quark models that in the case of massless-current quarks, chiral symmetry breaking usually triggers the generation of an anomalous magnetic for the quarks. In the same spirit, Singh [8] has also proven that, in theories in which chiral symmetry breaks dynamically, quarks can have a large a.m.m. On the other hand, Köpp et al [9] have provided a stringent bound on the a.m.m from high-precision measurements at LEP, SLC, and HERA. Given that a quark a.m.m likely exists, we include it in our formula. We show in the appendix how to get the anomalous part

$$\mu_{N\,anomalous} = \frac{1}{2}\sum_{i=u,d,s}\frac{Q_i}{2\overline{m}_i}a_i(\Delta_i - \Delta_{\bar{i}} + \frac{\overline{m}_i}{m_i}\delta i) \qquad \text{(Eq:4)}$$

The general formula giving the baryon magnetic moment in terms of $\Delta_i, \Delta_{\bar{i}}$ and $\delta i$ including the quark a.m.m has the form

$$\mu_N = \sum_{i=u,d,s}\frac{Q_i}{2\overline{m}_i}\left[\frac{1}{2(1+\frac{m_i}{\overline{m}_i})}(\Delta_i - \Delta_{\bar{i}} - \frac{\overline{m}_i}{m_i}\delta i) + (\frac{1+a_i}{2})(\Delta_i - \Delta_{\bar{i}} + \frac{\overline{m}_i}{m_i}\delta i)\right] \qquad \text{(Eq:5)}$$

Once the experimental quantities $\Delta_i, \Delta_{\bar{i}}, \delta i$ get measured to some accuracy, the above formula will serve to extract experimental values of the quark anomalous moment $a_i$, the effectives masses $\overline{m}_i$ and the ratio of the constituent quark mass to the quark effective mass $\frac{m_i}{\bar{m}_i}$. It is to note that due to the presence of the relativistic term the anomalous parameter could not be absorbed in a redefinition of the quark magnetic moment ( or equivalently in a change of the effective mass of the quark) . On the other hand the a.m.m $a_i$ are in general flavor dependant. This dependence is very model dependant but most of the calculations of the anomalous magnetic moments of the quark show that the flavor dependence is very small . To have an idea of the order of magnitude of the flavor dependence we inspect deviations of the ratios of the magnetic moments of the quarks $\frac{\mu_u}{\mu_d} \approx -2(1+a_u-a_d)$ and $\frac{\mu_s}{\mu_d} \approx \frac{m_d}{m_d}(1+a_s-a_d)$ from their SU(3) symmetry values respectively $-2$ and $\frac{m_d}{m_s}$. An estimation of these deviations [10] has already been given in the Chiral quark model of Georgi and Manohar. The anomalous quark moments were calculated to leading order by coupling photons to the quarks and Nambu-Goldstone bosons in one loop diagram. The outcome of these computations $\frac{\mu_u}{\mu_d} \approx -2+0.13$ and $\frac{\mu_s}{\mu_d} \approx \frac{m_d}{m_s}(1-0.07)$ show that the SU(3) breaking is very small .Other calculations in several quark models [6], in particular the second Nambu Jona-Lasinio considering only $u$ and $d$ quarks produces a even smaller deviation $\frac{\mu_u}{\mu_d} \approx -2+0.024$. Now that we have given the general formula relating electric and magnetic dipole moments to the hadron tensor charge, we may correct a wrong suggestion in the literature.The author [11] naively expects that if the quark a.m.m (e.d.m) is a tuneable parameter, the variation of the nucleon a.m.m (e.d.m) with respect to the quark(of flavour i) a.m.m (e.d.m) is equal to the tensor charge $\delta i(\delta i)$. These are a wrong suggestions. The correct answer is provided by equations (Eq:3) and (Eq:5).

$$\frac{\partial(\text{nucleon a.m.m})}{\partial(\text{quark}_i \text{ a.m.m})} = \frac{1}{2}(\Delta_i - \Delta_{\bar{i}} + \frac{\overline{m}}{m_i}\delta i)$$

$$\frac{\partial(\text{nucleon e.d.m})}{\partial(\text{quark}_i \text{ e.d.m})} = \delta_i + \delta_{\bar{i}}$$

III – Contributions to moments non expressible in terms of $\delta$

At this stage of the analysis it is necessary to look at the a.m.m and e.d.m separately. The magnetic and electric dipole-moment matrices $D$ of the quarks appear in the dimension-5 operator $\frac{i}{2}\overline{\psi}_L D\sigma_{\mu\nu}\psi_R F^{\mu\nu} + hc$, where the fermion fields are in the mass-eigenstate basis. Decomposing $D$ into hermitian and anti-hermitian parts $D = D_H + D_{AH}$ the dipole operator takes the form $\frac{1}{2}(\overline{\psi}D_H\sigma_{\mu\nu}\psi + \overline{\psi}D_{AH}\sigma_{\mu\nu}\gamma_5\psi)F^{\mu\nu}$ then by definition, the a.m.m and e.d.m of the quark are

$$d_i = -iD_{AH,ii}$$

$$a_i = \frac{m_i}{Q_i}D_{H,ii}$$

Although both moments have the property to flip the chirality of the quark (due to $\sigma_{\mu\nu}$), the electric dipole moment needs a T( factor i ) violating piece or equivalently CP violating piece (assuming CPT invariance). In the Standard Model CP non-conservation occurs both through a phase in the CKM matrix and in neutral as well as charged Higgs boson exchanges. CP odd neutron electric dipole moment vanishes at two loops. At three loops they have been estimated to be $d_n \approx 10^{-32\pm1} ecm$. These expectations are much lower than the current experimental limits $|d_n| \prec 6.3 \times 10^{-26} ecm$ [12]. There is also a CP-odd term the, $\theta$ term which is due to the topological charge of the QCD vacuum but this leads to an unnaturally small value $\theta \leq 10^{-10}$ [13]. Extensions of the standard model which bring additional sources of CP violations (supersymmetric models, extended technicolor models etc) are the appropriate place to estimate the electric dipole moment of the quark. . The m.m of the quark on the other hand is already appreciable within the standard model. We therefore analyse the non-additive contributions to the m.m within the standard model. Among possible potential contributions to the m.m of the baryon other than the quark intrinsic moments, we will consider only exchange magnetic moments for illustration. Exchange magnetic moments contribute a non-additive piece to the baryon magnetic moments. Non-additivity means this contribution will add an additional term to formula (Eq:5) not necessarily expressible in terms of $\delta i$. Our formula (Eq:5) is however safe for it has been shown in the chiral quark model [9], that this contribution is enough small to be neglected. The two- body exchange moments ( to consider only the leading) come from the exchange of one Nambu -Goldstone boson with one photon attached in all possible ways. The photon is coupled to the constituent quark and the meson through minimal substitution. A rough estimate of the size of the exchange moments gives the values $0.010 \mu_N$ which contribute a small amount to the baryon magnetic moments. This means that the leading contributions to the baryon magnetic moment come from the intrinsic magnetic moments of the quarks. The analogous formula for the electric dipole moment (Eq:3) on the contrary does not represent the leading contribution to the nucleon electric dipole moment. The triple gluon operator $c f_{\alpha\beta\gamma} G_{\alpha\mu\rho} G_{\beta\nu}{}^{\rho} G_{\gamma\lambda\sigma} \varepsilon^{\mu\nu\lambda\sigma}$ which contributes a non-additive piece to the e.d.m involves neither light-quark masses nor small mixing angles to make its contribution small. It is purely gluonic operator. Dimensional analysis may favor this operator with respect the to the light-quark electric dipole operators $\frac{i}{2}d_i\bar{\psi}\sigma_{\mu\nu}\gamma_5\psi F^{\mu\nu}$ which by chirality must contain at least one light quark mass factor. The chromoelectric dipole moment is another non-additive contribution which may be sizable. To summarize, all what we can say is that the e.d.m formula (Eq:5) is incomplete

$$d_N = \sum_{i=u,d,s} d_i(\delta_i + \delta_{\bar{i}}) + \cdots$$

with the ellipsis indicating eventually sizable non-additive pieces .We have seen that the quark e-d-m is sensitive to new physics beyond the standard model but ignorance of the complete formula make the exploration of new physics by means of the nucleon e.d.m quite impossible

IV – Conclusion

Electric and magnetic moments of the nucleon are static properties and hence associated with the nucleon at rest. The quark inside the nucleon are nevertheless strongly bound relativistic objects. Being relativistic the spin structure of the quarks will involve in general both the quark helicity distribution and the transversity. The latter encodes relativistic effects inside the nucleon. In this paper we proposed an expression relating the magnetic moment of the nucleon in terms of the quantities $\Delta_i - \Delta_{\bar{i}}$ and the tensor charge $\delta i$ and showed that this expression is valid up to negligible corrections from non additive potential contributions such as "exchange moments". An analogous expression for the electric dipole moment is expressed in term the combination $\delta_i + \delta_{\bar{i}}$ but eventually sizable contributions from CP violating sources not necessary associated intrinsically to the quarks make this incomplete expression restricted use.

Appendix:

To prove formula (Eq:4), we Fourier transform the anomalous part of the magnetic moment operator. We consider only one flavor and no antiquark to simplify notations

$$\bar{\mu}_N\big|_{anomalous} = -\frac{aQ}{2}\left\langle P\uparrow\left|\int \frac{\partial}{\partial \vec{q}}\times(\bar{\psi}(k')\frac{\vec{\sigma}^v}{2m}\psi(k)q_v)\frac{d^3p}{(2\pi)^3}\right|_{q=0}\right|P\uparrow\right\rangle$$

with $\vec{q} = \vec{k} - \vec{k}'$, $\vec{p} = \dfrac{\vec{k}+\vec{k}'}{2}$ and $\vec{\sigma}^v$ is a vector whose components are $\sigma^{iv}$. Then write

$$-\sigma^{iv}q_v = -(\vec{q}\times\vec{\Sigma})^i + i\vec{\alpha}q_0$$
$$\sigma^{ij} = \epsilon^{ijk}\Sigma_k$$
$$\sigma^{io} = -i\alpha^i$$

Differentiate each term of the above expression

$$i\frac{\partial}{\partial \vec{q}}\times(\bar{\psi}\vec{\alpha}q_0\psi)\bigg|_{q=0} = i\frac{\vec{k}\times\bar{\psi}\vec{\alpha}\psi}{k^0}$$

$$= \frac{m}{k_0}\bar{\psi}\vec{\gamma}\gamma_5\psi - \bar{\psi}\vec{\gamma}\gamma_5\gamma_0\psi$$

$$= \frac{m}{k_0}\psi^\dagger\vec{\Sigma}\psi - \bar{\psi}\vec{\Sigma}\psi$$

$$-\frac{\partial}{\partial \vec{q}}\times(\vec{q}\times\bar{\psi}\vec{\Sigma}\psi)\bigg|_{q=0} = 2\bar{\psi}\vec{\Sigma}\psi$$

To get the last term in the first equation we used the identities

$$(\vec{\gamma}.\vec{k})\vec{\gamma} = -\vec{k} - i(\vec{\gamma}\times\vec{k})\gamma_0\gamma_5$$
$$\vec{\gamma}(\vec{\gamma}.\vec{k}) = -\vec{k} + i(\vec{\gamma}\times\vec{k})\gamma_0\gamma_5$$

Using the definition of the tensor and axial currents

$$\langle PS | \int dx^3 \psi_i^\dagger \frac{\vec{\Sigma}}{2} \psi_i | PS \rangle = \Delta_i \vec{S}$$

$$\langle PS | \int dx^3 \overline{\psi}_i \vec{\Sigma} \psi_i | PS \rangle = \vec{\delta}_i$$

we get for the anomalous part

$$\mu_N \big|_{anomalous} = \frac{aQ}{4m} \langle P\uparrow | \int \frac{m}{k_0} \psi^\dagger \vec{\Sigma} \psi + \overline{\psi} \vec{\Sigma} \psi ) \frac{d^3k}{(2\pi)^3} | P\uparrow \rangle$$

$$= aQ \langle P\uparrow | \frac{\vec{S}}{2\overline{m}} + \frac{1}{2m}(\frac{1}{2}\vec{\delta}) | P\uparrow \rangle$$

$$= \frac{aQ}{4\overline{m}} (\Delta + \frac{\overline{m}}{m} \delta)$$